\documentclass[11pt,twoside]{article}
\usepackage{asp2014}

\aspSuppressVolSlug
\resetcounters
\newcommand{\teff}{$T_{\rm eff}$}
\newcommand{\Msun}{$M_{\rm \odot}$}
\begin{document}

\title{HD188112: Supernova Ia progenitor?}
\author{M. Latour,$^1$ A. Irrgang,$^1$, U. Heber$^1$ and V. Schaffenroth$^{1,2}$}
\affil{$^1$Dr. Karl Remeis-Observatory \& ECAP, Astronomical Institute,
Friedrich-Alexander University Erlangen-Nuremberg, Bamberg, Germany; \email{marilyn.latour@fau.de}}
\affil{$^2$Institute for Astro- and Particle Physics, University of Innsbruck, Innsbruck,  Austria }

\paperauthor{M. Latour}{marilyn.latour@fau.de}{}{Friedrich-Alexander University Erlangen-Nuremberg}{Dr. Karl Remeis-Observatory \& ECAP, Astronomical Institute}{Bamberg}{}{96049}{Germany}
\paperauthor{A. Irrgang}{}{}{Friedrich-Alexander University Erlangen-Nuremberg}{Dr. Karl Remeis-Observatory \& ECAP, Astronomical Institute}{Bamberg}{}{96049}{Germany}
\paperauthor{U. Heber}{}{}{Friedrich-Alexander University Erlangen-Nuremberg}{Dr. Karl Remeis-Observatory \& ECAP, Astronomical Institute}{Bamberg}{}{96049}{Germany}
\paperauthor{V. Schaffenroth}{}{}{Friedrich-Alexander University Erlangen-Nuremberg}{Dr. Karl Remeis-Observatory \& ECAP, Astronomical Institute}{Bamberg}{}{96049}{Germany}

\begin{abstract}
HD188112 is an extremely low mass white dwarf in a close binary system. According to a previous study, the mass of HD188112 is $\sim$0.24 \Msun\ and a lower limit of 0.73  \Msun\ could be put for the mass of its unseen companion, a compact degenarate object. We used HST STIS spectra to measure the rotational broadening of UV metallic lines in HD188112, in order to put tighter constraints on the mass of its companion. By assuming that the system in is synchronous rotation, we derive a companion mass between 1.05 and 1.25  \Msun.  We also measure abundances for magnesium, silicon, and iron, respectively log $N$(X)/$N$(H) = $-$6.40, $-$7.25, and $-$5.81. The radial velocities measured from the UV spectra are found to be in very good agreement with the prediction based on the orbital parameters derived in the previous study made a decade ago. 

\end{abstract}

\section{Introduction}
HD 188112 is a bright ($V$=10.2) and nearby star ($\sim$80 pc, \textsc{Hipparcos} parallax) that is found below the Extreme Horizontal Branch in the log $g$ - \teff\ plane (21 500 K, log $g$ = 5.6). The spectroscopic parameters, combined with the parallax distance allowed to determine a mass of 0.24 $^{+0.10}_{-0.07}$ \Msun\ for the star, too low to sustain helium burning, thus making it an extremely low mass (ELM) white dwarf (WD) \citep{heb03}. This mass estimate is also in agreement with the theoretical evolutionary tracks for post red giant branch of \citet{dri98}. HD 188112 is part of a close binary system with an orbital period of $\sim$14.5 hr and a semi-amplitude of K = 188 km s$^{-1}$, which lead to a lower limit of 0.73 \Msun\ for the companion's mass. A main sequence star of such a mass would contribute to the infrared flux of the system, but no near infrared excess is seen in the 2MASS data. There is also no trace of the companion in the high-resolution optical data ($\lambda/\Delta\lambda$ = 48,000), thus indicating that this companion is a rather massive WD \citep{heb03}. Without knowing the inclination of the system, it is not possible to put further constraints on the mass of the unseen companion. However, assuming tidally locked rotation, the inclination of the system can be determined by measuring the rotational broadening (v$_{rot}$sin$i$) of metal lines. Given the fact that the optical spectra is devoid of metal lines, except for one Mg~\textsc{ii} line too weak for such a measurement, UV spectra were needed in order to determine the rotational broadening. 

\section{Observations}

We obtained Hubble Space Telescope (HST) high resolution ($\lambda/\Delta\lambda$ = 114,000) NUV (2660 $-$ 2935 \AA) and FUV (1242 $-$ 1440 \AA) STIS spectra of HD 188112. The FUV observations correspond to 22 individual exposures of 120 s that were then corrected for radial velocity shifts and co-added. There are also two NUV exposures (1040.2 and 2860.2 s), taken in Time-Tag mode, that were afterwards divided into shorter exposures, corrected for radial velocity shifts and co-added in order to have two spectra with sharp (unblurred) metallic lines.
As expected from the rather plain optical spectrum of the star, the UV range is not crowded with lines. The NUV spectrum features a set of Mg~\textsc{ii} lines as well as a few Fe~\textsc{ii} lines. The FUV spectrum shows a lot more lines, of Si~\textsc{ii-iii-iv}, S~\textsc{ii}, Fe~\textsc{iii} as well as additional ones that still need to be identified. Even if this portion of the spectrum is richer in lines, the majority of them are unblended and well isolated, thus suitable for abundances and vsin$i$ measurements. 

\articlefigure{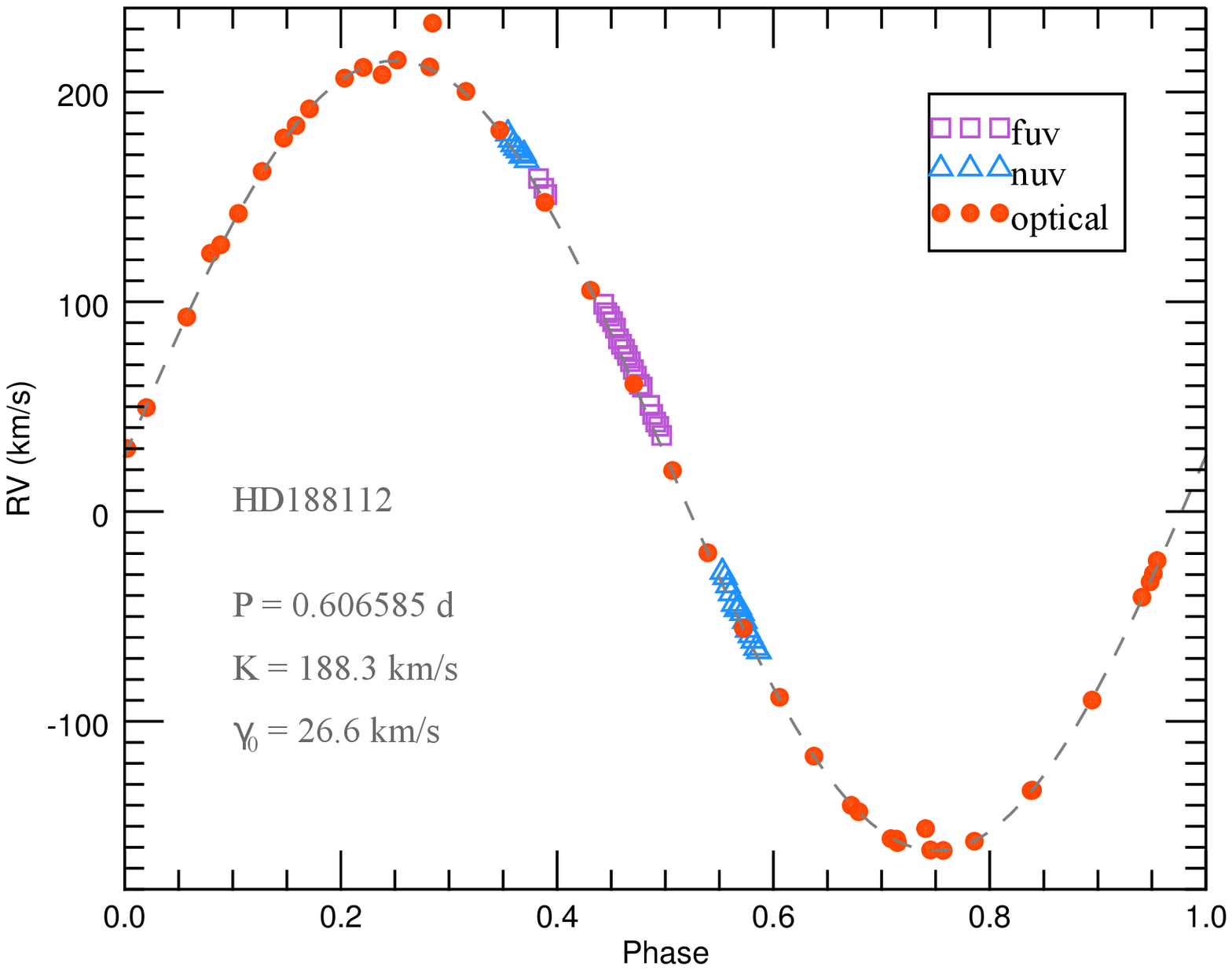}{fig1}{Radial velocity curve of HD 188112 phased over the parameters detemined in \citet{heb03}. The radial velocities measured with the FUV and NUV spectra are plotted in addition to the optical ones. }

\section{Analysis}

\subsection{Radial velocity measurements}

Radial velocities were measured for the 22 FUV spectra using Si~\textsc{iii} and~\textsc{iv} lines while the Mg~\textsc{ii} lines were used in the 29 NUV exposures made out of the Time-Tag data. Fig. \ref{fig1} shows the phased radial velocity curve, according to the parameters of \citet{heb03}. Our additional UV velocities end up very close to the predicted ones, meaning that the orbital parameters are well constrained with the optical data.

\subsection{Spectral analysis}

Our model atmospheres and synthetic spectra are computed using the hybrid, non-LTE approach discussed in \citet{prz06} and \citet{nie07,nie08}. The atmospheric structure, such as the temperature and density stratification, is computed with \textsc{Atlas12} \citep{kur96}. This atmospheric structure is then used by two additional codes that compute the non-LTE population numbers (\textsc{Detail}) and the final synthetic spectra with detailed line-broadening (\textsc{Surface}). 

\articlefigure{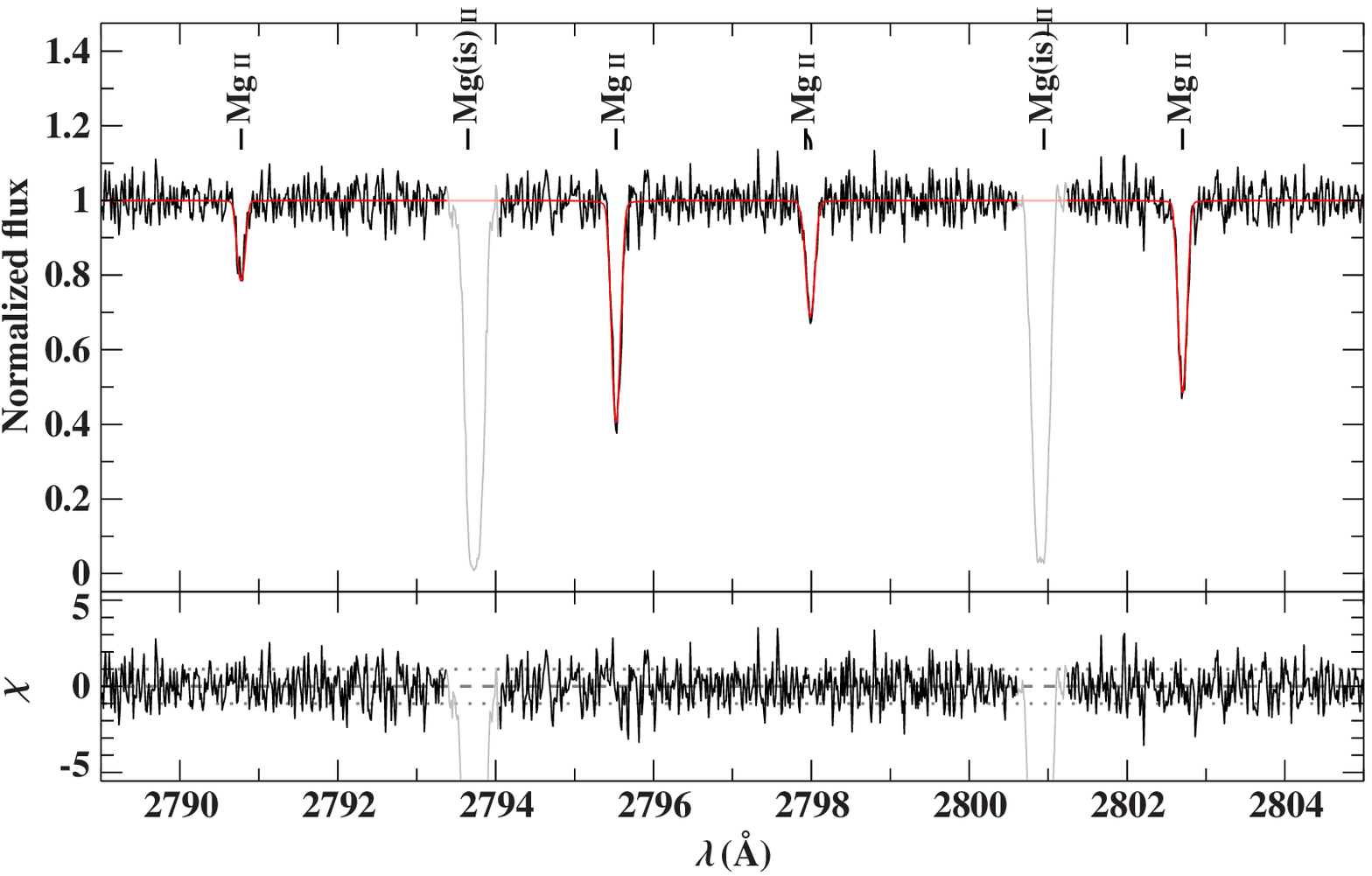}{fig2}{Resulting fit of the Mg lines (and v$_{rot}$sin$i$). The interstellar components (grey) were excluded from the fitting procedure. }

We computed small metal grids of synthetic spectra for various abundances of magnesium, silicon and iron, keeping the fundamental parameters fixed at the values determined by \citet{heb03}.
In a first step, we used the Mg and Si lines to measure the rotational velocity (v$_{rot}$sin$i$), as well as their abundances. Resulting fits for the Mg lines and three Si~\textsc{iii} lines are shown in Fig. \ref{fig2} and \ref{fig3}. A dozen or so silicon lines were simultaneously fitted, and we obtained a very good fit for the three different ionization stages (Si~\textsc{ii-iii-iv}). 
Our resulting parameters are : 
\begin{itemize}
\item log $N$(Mg)/$N$(H) = $-$6.40 $\pm$ 0.07 ($\sim$ 1/100 solar)
\item log $N$(Si)/$N$(H) = $-$7.25 $\pm$ 0.05 ($\sim$ 1/560 solar)
\item v$_{rot}$sin$i$ = 8.2 $\pm$ 0.2 km s$^{-1}$
\end{itemize}
with the uncertainties derived while re-doing the fitting procedure with various \teff\ and log $g$ within the error limits given by \citet{heb03} ($\pm$ 500 K and $\pm$ 0.05 dex).

\articlefigure{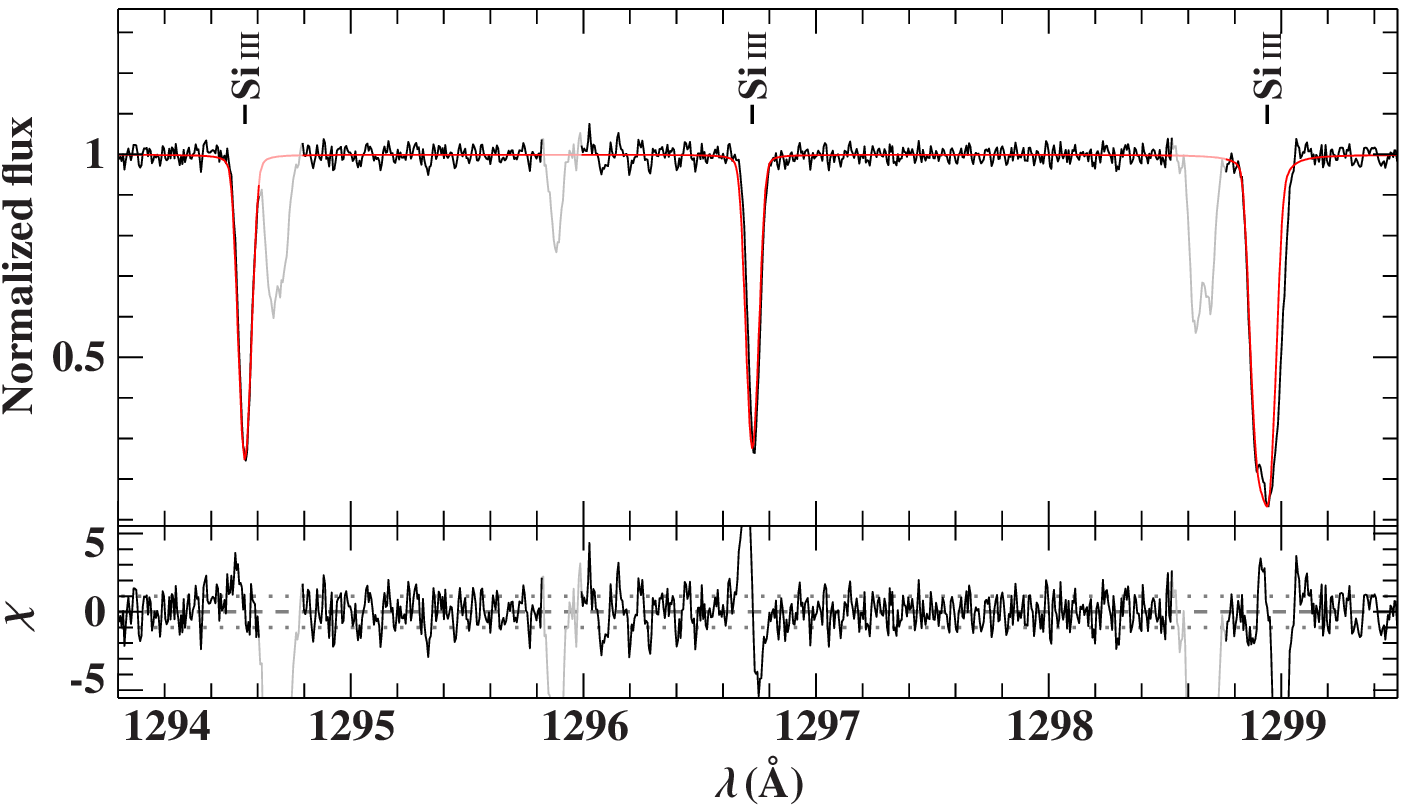}{fig3}{Resulting fit for three of the Si lines (and v$_{rot}$sin$i$).}

We also fitted the iron lines present in the FUV and NUV spectra, this time keeping the v$_{rot}$sin$i$ value to the one stated above (see Fig. \ref{fig4}). We found a small discrepancy between the abundances indicated by the Fe~\textsc{iii} and~\textsc{ii} lines: 
\begin{itemize}
\item log $N$(Fe~\textsc{iii})/$N$(H) = $-$5.81 $\pm$ 0.04 ($\sim$ 1/20 solar)
\item log $N$(Fe~\textsc{ii})/$N$(H) = $-$5.62 $\pm$ 0.08
\end{itemize}

\articlefigure{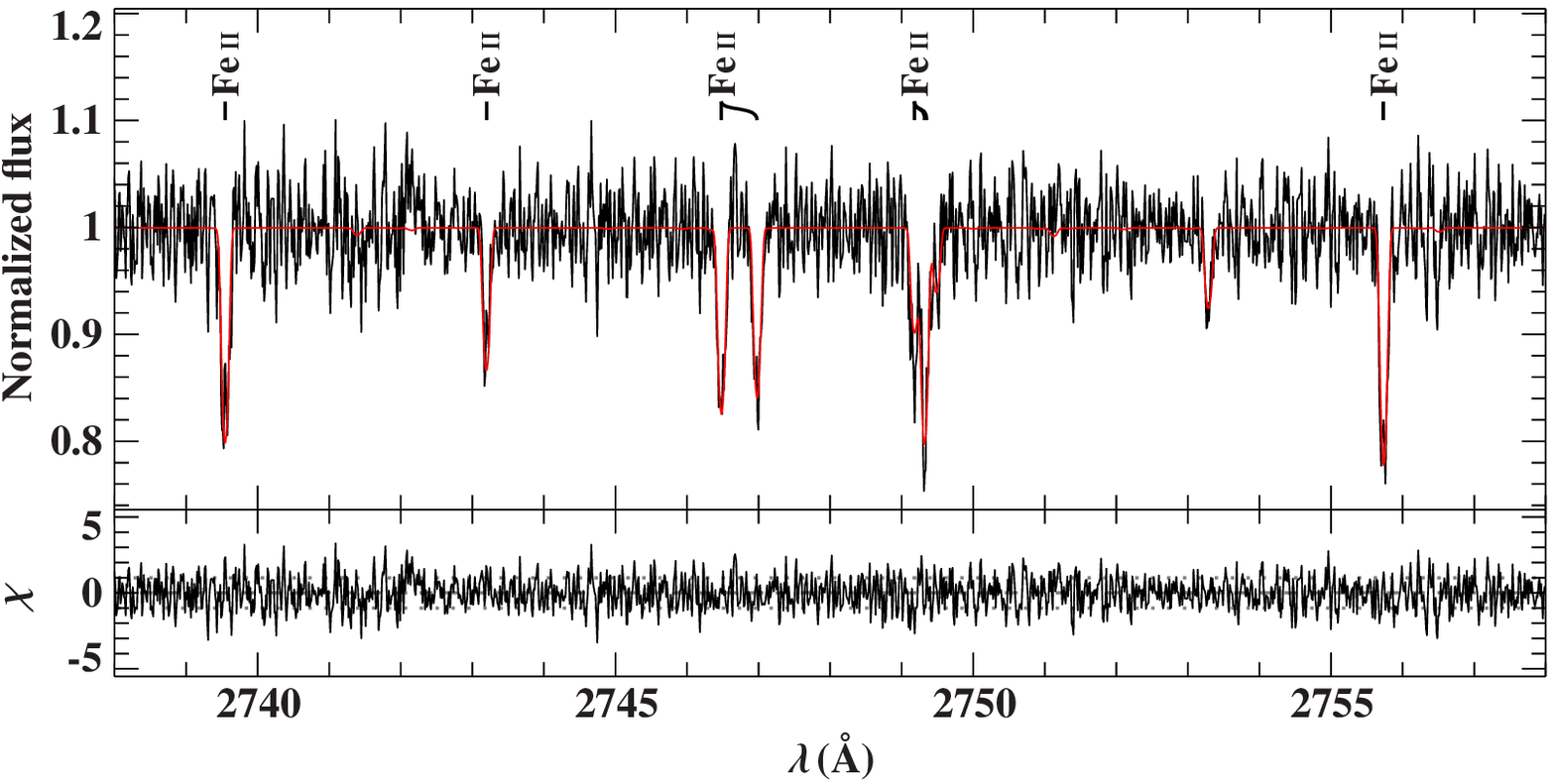}{fig4}{Resulting fit for the Fe~\textsc{ii} lines in the NUV spectrum.}

\section{Conclusion}

By making the assumption that the system is tidally locked, so that the rotation period of HD188112 is the same as its orbital period, then the vsin$i$ measured (8.2 $\pm$ 0.2 km s$^{-1}$) and the mass of HD188112 (0.24 $\pm$ 0.1 \Msun) allow to derive a mass between 1.05 and 1.25 \Msun\ for the unseen WD companion. However we cannot currently ascertain that the system is in synchronous rotation. 

Nevertheless, if the mass of the companion is below $\sim$1.10 \Msun\ and its core composed of carbon and oxygen, the system could be a candidate for the double detonation scenario \citep{fink10}. According to this scenario, a surface detonation in the helium shell of the massive WD companion would trigger the detonation and explosion of its carbone/oxygen core, leading to a supernova Ia event. HD188112 could play the role of the helium donor. 
On the other hand, if the companion has a mass above $\sim$1.16 \Msun, then the total mass of the system would be above the Chandrasekhar limit.

Additional abundances and upper limits still have to be measured for other metallic elements. No lines of carbon, nitrogen or oxygen are visible in the UV spectra, so only upper limits will be derive for these elements, but we do see some sulphur lines in the FUV spectra and some nickel lines are also expected to be seen. 
Up to now, we found a rather large abundance variation among Mg, Si and Fe. We will also try to put constraints on the microturbulent velocity.
New radial velocities will also be obtained from recent optical spectra and combined with the UV ones we could derived new orbital parameters for HD188112, though we don't expect those parameters to change much, given the well predicted values for our UV radial velocities.

\acknowledgements M. L. acknowledges funding by the Deutsches Zentrum f\"{u}r Luft- 
und Raumfahrt (grant 50 OR 1315).



\end{document}